\documentstyle[12pt]{article}

\thicklines                         % thick straight lines for pictures (LaTeX)
\def\a{\alpha}
\def\b{\beta}

\def\e{\epsilon}                % Also, \varepsilon
\def\j{\psi}

\def\m{\mu}

\def\o{\omega}
\def\th{\theta}                  %       \vartheta
\def\s{\sigma}                  %       \varsigma

\def\car{{\cal R}}

\def\rhs{\mbox{r.h.s.} }

\def\ie{\mbox{i.e.} }
\def\eg{\mbox{e.g.} }

\def\APH#1{Ann. Phys. {\bf #1}}

\def\NPB#1{Nucl. Phys. {\bf B#1}}
\def\NPBP#1{Nucl. Phys. (Proc. Suppl.) {\bf B#1}}
\def\PLB#1{Phys. Lett. {\bf B#1}}
\def\PRD#1{Phys. Rev. {\bf D#1}}
\def\PR#1{Phys. Rev. {\bf #1}}

\def\APP#1{Acta. Phys. Pol. {\bf B#1}}

\def\Im{{\rm Im\,}}

\def\sbra#1{\left\langle #1\right|}             % variable < |
\def\sket#1{\left| #1\right\rangle}             % variable | >

\begin{document}
\hfill WIS--93/56--JUNE--PH
\par
\begin{center}
\vspace{15mm}
{\large\bf Constraints on the Existence of Chiral Fermions\\
in Interacting Lattice Theories}\\[5mm]
{\it by}\\[5mm]
Yigal Shamir\\
Department of Physics\\
Weizmann Institute of Science, Rehovot 76100, ISRAEL\\
email: ftshamir@weizmann.bitnet\\[15mm]
{ABSTRACT}\\[2mm]
\end{center}
\begin{quotation}
  It is shown that an interacting theory, defined on a regular lattice,
must have a vector-like spectrum if the following
conditions are satisfied: (a)~locality, (b)~relativistic
continuum limit without massless bosons, and (c)~pole-free effective
vertex functions for conserved currents.

  The proof exploits the zero frequency inverse retarded propagator of
an appropriate set of interpolating fields as an effective quadratic
hamiltonian, to which the Nielsen-Ninomiya theorem is applied.

\vspace{3ex}
\end{quotation}

\newpage

  The only rigorous way, presently known to us, to define non-abelian
gauge theories, relies on the lattice as a regulator. The observed
fermion spectrum fits into a chiral representation of
SU(3)$\times$SU(2)$\times$U(1), and so the construction of a consistent
chiral gauge theory on the lattice has been a major goal in theoretical
physics.

  In spite of extensive efforts, this program
has been unsuccessful to date. The basic stumbling block is
the doubling problem~[1]. A naive discretization
of the continuum hamiltonian of a Weyl fermion gives rise to eight
Weyl fermions in the classical continuum limit of the lattice
hamiltonian. If one starts with a Dirac fermion, the doublers can be
eliminated by introducing the Wilson term. But the price is that the
axial symmetry of the classical continuum hamiltonian is lost.

  Following the work of Karsten and Smit~[2],
the precise conditions for the presence of doublers in a free
fermionic theory defined on a regular lattice were stated by Nielsen and
Ninomiya  as a no-go theorem~[3]. They assume the existence of a set of
exactly conserved, locally defined charges which admit discrete
eigenvalues.
The Nielsen-Ninomiya theorem then asserts that there must be an equal
number of positive helicity and negative helicity fermions in every
complex representation of the symmetry group, provided the Fourier
transform of the free hamiltonian
has a continuous first derivative. (Recall
that chirality equal helicity for massless fermions). The
Nielsen-Ninomiya theorem applies in particular when the hamiltonian has
a short range, and the charges are constructed canonically and generate
a compact Lie group.

  The absence of chiral fermions is essentially a counting theorem
about the zeros of the free hamiltonian in the Brillouin zone.
   In three space dimensions, a massless fermion
corresponds to level crossing which is described by the
effective two-by-two hamiltonian
$$
H_{eff}({\bf p})=\pm{\bf \s}\cdot ({\bf p}-{\bf p}_c) +
O(({\bf p}-{\bf p}_c)^2)\,.
\eqno(1)
$$
The point ${\bf p}_c$ is called a {\it degeneracy point}. The $\pm$
signs correspond to the helicity of the fermion.
The Nielsen-Ninomiya theorem is then a consequence of  theorems
in algebraic topology which exploit the fact that the Brillouin zone is
topologically a three-torus.

  The continuum limit of asymptotically free gauge theories is
achieved at vanishing bare coupling. Together with the success of
perturbative QCD in deep inelastic scattering, this leads to the
generally accepted view that the fermionic spectrum can be correctly
determined by setting the gauge couplings to zero. In the absence of
other interactions, the Nielsen-Ninomiya theorem implies that the
fermionic spectrum must be vector-like provided the
hamiltonian has a short range.

  Attempts to avoid fermion doubling by using long range lattice
derivatives lead to various inconsistencies at the level of weak
coupling perturbation theory~[4]. Important examples include the
SLAC derivative  which avoids the
extra zeros by creating a discontinuity in the dispersion relation, and
a method due to Rebbi which is characterized by the presence
of a pole in the dispersion relation. The former suffers from
Lorentz violations and non-locality~[5] while the latter suffers from
the presence of ghosts~[6].

  As it stands, the Nielsen-Ninomiya theorem does not apply if the
lattice model contains some strong interactions. This observation have
led to several proposals~[7] for constructing chiral gauge theories
on the lattice which exploit a common strategy. One starts with a model
containing only fermions and (possibly) scalar fields. In addition to
standard quadratic terms, one introduces judiciously chosen strong
interactions among these fields which are operative at the lattice
scale, and vanish in the classical continuum limit.
Local symmetries of the desired continuum theory -- the target theory --
should appear at this stage as exact global symmetries of the model.
One then tries to find a point in the phase diagram in which
all the doublers decouple.
It is crucial that, at the same time, the to-be-gauged
global symmetries are not broken spontaneously.

  If this program were successful, a consistent chiral
gauge theory could be obtained by turning on the gauge interactions in
such a model. However, explicit model calculations have lead
to negative conclusions in all cases studied so far~[8].

  Our purpose in the present letter is to provide a general treatment of
the problem, which applies for example also to similar models
based on the recently proposed domain wall fermions~[9].
We will prove that under mild assumptions, which are directly related
to the physical properties of a consistent continuum limit,
the spectrum is necessarily vector-like. In this letter we report our main
results, whereas the full details will appear elsewhere~[10].

  We will consider a hamiltonian defined on a regular space lattice
(time is continuous)
that has a compact global symmetry group which is not spontaneously
broken. The conserved generators of the global symmetries are assumed
to be the sum over all lattice points of a local density.
  The idea is to use the zero frequency inverse retarded propagator of an
appropriate set of interpolating fields, denoted $\car^{-1}({\bf p})$,
as an effective quadratic hamiltonian which satisfies all the
assumptions of the Nielsen-Ninomiya theorem.

  The crucial element of our theorem is the demonstration that
a sufficiently local anti-commutator,
of the kind expected in theories with a short range hamiltonian,
gives rise to an {\it analytic} $\car({\bf p})$. The only singularities
occur at {\it generalized degeneracy points}, which are  those
points in the Brillouin zone where the hamiltonian admits
eigenstates of vanishing energy. The proof of analyticity invokes the
``edge of the wedge'' theorem~[11], and it is an adaptation
to the lattice context of classic results from the theory of
dispersion relations.

\vspace{1ex}

 In more detail, if a particle can be created in a  causal
process, there should exist a {\it local interpolating field} which
has a finite probability to create the particle by acting on the vacuum.
The particle should then generate a singularity in the two point function
of the interpolating field.

   The two point function we will consider is the {\it retarded
anti-commutator}. Suppressing possible colour and
flavour indices it is defined by
$$
R_{\a\b}({\bf x},t)=i\th(t)\,\sbra{0}\{\j_\a({\bf x},t)\,,
\j^\dagger_\b({\bf 0},0)\}\sket{0}\,.
\eqno(2)
$$
We also introduce the space and space-time Fourier transforms
$$
\hat{R}_{\a\b}({\bf p},t)=\sum_{\bf x}
e^{ - i{\bf p \cdot x}}\,R_{\a\b}({\bf x},t) \,,
\eqno(3)
$$
$$
\tilde{R}_{\a\b}({\bf p},\o)=\int_0^\infty dt\, e^{i\o t}
\hat{R}_{\a\b}({\bf p},t)\,,
\eqno(4)
$$
and define
$$
\car_{\a\b}({\bf p})=\lim_{\e\to 0}\tilde{R}_{\a\b}({\bf p},\o=i\e) \,.
\eqno(5)
$$

  We first observe that $R({\bf x},t)$ is {\it bounded}.
This is a trivial consequence of translation invariance and
the fact that $\j_\a({\bf 0},0)$ is a well defined operator on
the Hilbert space. Thus, there exists $0<b_1<\infty$ such that
$$
|R_{\a\b}({\bf x},t)| \le
\left\Vert
\stackrel{ }{\j}\!
\sket{0}\right\Vert^2 +
\left\Vert
\j^\dagger\sket{0}\right\Vert^2
\le b_1 \,.
\eqno(6)
$$

  For similar reasons, at fixed ${\bf x}$,
$R({\bf x},t)$ is an analytic function of $t$.
As a result, $R({\bf x},t)$ cannot vanish identically outside the light
cone, for then it would be zero everywhere.

  We should therefore discuss the rate at
which $R({\bf x},t)$ tends to zero at large space-like separations.
We will say that $R({\bf x},t)$ is {\it local} if it can be bounded
by an exponential, \ie if there are positive constants $c$, $b_2$
and $\m$ such that for all $|{\bf x}|>ct$
$$
|R_{\a\b}({\bf x},t)| \le
\min \left\{b_1, b_2 e^{-\m(|{\bf x}|-ct)}\right\} \,.
\eqno(7)
$$
We will say that $R({\bf x},t)$ is {\it strongly local} if it decreases
faster than an exponential, \ie if it satisfies a bound of
the form~(7) for every $\m$.

  We will assume below that $R({\bf x},t)$ is local or strongly local.
This will allow us to prove the analyticity of $\car({\bf p})$.
In view of the intimate relation between causality and analyticity in
the continuum,  an exponentially bounded anti-commutator on the lattice
should be a necessary condition for causality in the continuum limit.
We comment, however, that to exclude the existence of chiral fermions,
all that is needed is that $\car^{-1}({\bf p})$ have a continuous first
derivative. Consequently, it is sufficient to assume a much weaker form
of locality, namely, that the anti-commutator is bounded by an
appropriate inverse power of ${\bf x}$~[10].

  The last thing we need is the notion of a {\it generalized degeneracy
point}. Introducing the {\it advanced} anti-commutator
$\tilde{A}({\bf p},\o)$ we define for real values of ${\bf p}$ and $\o$
$$
E_0({\bf p})=\sup\{\o\,|\,\tilde{R}({\bf p},\o')=
\tilde{A}({\bf p},\o')\mbox{  if  } |\o'|<\o\}\,.
\eqno(8)
$$
The physical meaning of this definition is that $E_0({\bf p})$ is the
lowest possible energy for eigenstates with momentum ${\bf p}$.
We define a generalized degeneracy point by the condition
$E_0({\bf p}_c)= 0$.
Thus, ${\bf p}_c$ is a generalized degeneracy point
if it is the end point of a gap-less continuous spectrum.

\vspace{2ex}

\noindent {\it Lemma}. Assume that $R(x,t)$ is local in the sense
of eq.~(7). Then (a) $\tilde{R}({\bf p},\o)$ is holomorphic in the
domain $\Im\o>0$, $|\Im{\bf p}|<\min\{c^{-1}\Im\o\,,\m\}$; (b)
$\car({\bf p})$ is analytic with singularities only at generalized
degeneracy points.

\vspace{1ex}

\noindent {\it Proof}. By assumption, the \rhs of eq.~(3) is bounded
by the \rhs of eq.~(7) times $e^{|{\bf x}|\,|\Im{\bf p}|}$.
Hence, the sum in eq.~(3) converges absolutely and
$\hat{R}_{\a\b}({\bf p},t)$ is holomorphic in the domain
$|\Im{\bf p}|<\m$.
Morever, $\hat{R}_{\a\b}({\bf p},t)$ can be bounded by a polynomial of
third degree in $t$ times $e^{ct\,|\Im{\bf p}|}$.
The presence of the damping factor
$e^{-t\,\Im \o}$ then implies that the integral on the \rhs of eq.~(4)
converges absolutely for $|\Im{\bf p}|<\min\{c^{-1}\Im\o\,,\m\}$.
This proves (a).  Notice that if  $R(x,t)$ is strongly local
than $\hat{R}_{\a\b}({\bf p},t)$ is an entire function of ${\bf p}$.
In this case $\tilde{R}({\bf p},\o)$ is holomorphic in the forward cone
$|\Im{\bf p}|<c^{-1}\Im\o$.

  In order to prove (b) we notice that the {\it advanced} anti-commutator
has similar properties except that the sign of
$\Im\o$ is now negative.
A straightforward application of the edge of the wedge theorem~[11]
now implies that the common boundary function $\car({\bf p})$ is
analytic, with singularities only at generalized degeneracy points.
This proves~(b).

\vspace{1ex}

  The analyticity of $\car_{\a\b}({\bf p})$ away from generalized
degeneracy points, implies that there can be no obstructions to the
smooth motion throughout the Brillouin zone from one zero of
$\car_{\a\b}^{-1}({\bf p})$ to another, {\it provided} we exclude the
possibility of {\it poles} in $\car_{\a\b}^{-1}({\bf p})$.
We forbid this situation by
{\it assuming} that the elements of the matrix
$\car_{\a\b}^{-1}({\bf p})$ are bounded.
We comment that, together with the assumption that symmetries
are not broken spontaneously,  this is
equivalent via the Ward identities to the
requirement that effective vertex functions, defined as
correlation functions of conserved currents and the interpolating
fields, are pole free.

  The presence of a pole in $\car_{\a\b}^{-1}({\bf p})$
may reflect a kinematical
singularity, arising from a bad choice of interpolating fields.
It may happen if (a) one uses two interpolating fields for two fermions
in different corners of the Brillouin zone,  whereas actually they can
both be interpolated by a single field, or (b) if not all fermions are
interpolated. In both cases it is possible to identify the kinematical
nature of the singularity, and to construct a new, {\it admissible}
set of interpolating fields which is free of this problem.
Details will be given elsewhere~[10].

  If a pole in $\car_{\a\b}^{-1}({\bf p})$
is not an artifact of an inadmissible set
of interpolating fields, it cannot arise unless the
hamiltonian is highly non-local.
In a very general context, it has been shown that such poles give
rise to the appearance of ghosts in one loop diagrams once gauge
fields are introduced~[6]. The reason is that, via the Ward identity,
such poles appear in the vertex function, but they contribute to
the vacuum polarization with the wrong sign.
Interpreted in a hamiltonian language, this
result implies that the action of a local current on the vacuum takes
one outside the Hilbert space, which is unacceptable. We thus expect
that it should be possible to extend our theorem and to rigorously
exclude the presence of such poles in a consistent quantum theory.

  By construction, $\car^{-1}({\bf p})$ is a hermitian matrix.
In order to show that it qualifies as an effective hamiltonian which
satisfies the assumptions of the Nielsen-Ninomiya theorem,
what is left for us to do is to
show that it has a continuous first derivative at
generalized degeneracy points and that its zeros can be
identified with massless fermions.

  In order to characterize the singularities of
$\car^{-1}({\bf p})$, we now assume that at sufficiently
large distances physics is correctly described by an effective
lagrangian of massless fermions interaction only via non-renormalizable
couplings. The justification of this assumption is that, in all models
which attempt to decouple the doublers dynamically, the aim is to
achieve a continuum limit with the above properties (before the gauge
interactions are turned on). We  demand
that all singularities of $\car^{-1}({\bf p})$ should be compatible
with those allowed by the effective lagrangian.

  Let us denote by ${\bf p}_{phys}$ the momentum variable which
transforms homogeneously under the Lorentz group in the low energy limit.
The only singularity compatible with the assumed form of the low energy
effective lagrangian is
$$
\pm{1\over {\bf \s\cdot p}_{phys}
\left(1+O({\bf p}_{phys}^4 \log{\bf p}_{phys}^2)\right) }\,.
\eqno(9)
$$
We have ${\bf p}_{phys}^4$ in front of the logarithmic term
because this term involves at least two powers of coupling constants,
and all coupling constants have a negative mass dimension which is
at least two.

  Clearly, an allowed singularity of $\car({\bf p})$ can be
obtained by substituting
${\bf p}-{\bf p}_c$ instead of ${\bf p}_{phys}$ in eq.~(9). We call
such a singularity a {\it primary singularity}. The precise definition
is as follows. A generalized degeneracy point ${\bf p}_c$ is a
primary singularity if there exists a unitary transformation $U$
such that
$$
\lim_{ {\bf p}\to{\bf p}_c}
\left[ {\bf \s\cdot}({\bf p}-{\bf p}_c)\otimes I_1 \right]
U \car({\bf p}) U^\dagger = \left[ I \otimes A \right]\,.
\eqno(10)
$$
In eq.~(10), $I$ is the identity matrix in spin space, $I_1$ is the
identity matrix in colour and flavour space, and $A$ is a diagonal
matrix $A=diag(Z_1,\ldots,Z_k,0,\ldots,0)$. The $Z$-s are non-zero
constants, which are in one-to-one correspondence with massless
fermions. The helicity of the massless fermion is determined by the
sign of the corresponding $Z$.

 In addition, there will in general
be  {\it secondary singularities} at points which are
integer multiples of the primary singularities. These points
correspond to multi-particle spectra
having the same quantum numbers as the original particle.
The leading contribution of the gap-less spectrum to
$\car({\bf p})$ at a secondary singularity point can take the form
$$
\pm {\bf \s\cdot p}_{phys}\,{\bf p}_{phys}^2 \log{\bf p}_{phys}^2\,.
\eqno(11)
$$
As before, ${\bf p}_{phys}={\bf p}-{\bf p}_c$.
This form is dictated by the requirement that, once we sum
$\car({\bf p})$ over all points that correspond to the same
${\bf p}_{phys}$, we will obtain an expression compatible with the
expansion of the denominator of eq.~(9) in powers of the coupling
constant. Notice that there is no reason that $\car({\bf p})$ should
vanish at a secondary singularity point, because it always receives
additional, regular contributions from finite energy branches of the
spectrum.

  We now collect all our intermediate results together in the following
theorem.

\vspace{2ex}

\noindent {\it Theorem}. Consider a hamiltonian defined on a regular
lattice. Assume the existence of a compact global symmetry group which
is not spontaneously broken. Assume also that the continuum limit is
relativistic and that the only massless particles are fermions.
Under these assumptions, there is an equal number of left handed and
right handed fermions in every complex representation of the global
symmetry group, provided $R({\bf x},t)$ is local and
$\car^{-1}({\bf p})$ is bounded for every admissible set of
interpolating fields.

\vspace{1ex}

\noindent {\it Proof}. Consider all sets of interpolating fields which
satisfy the above assumptions and which belong to a given complex
representation. Choose a maximal set. By this we mean that the total
number of $Z$-s, as determined by the limiting procedure~(10) and summed
over all primary singularity points, is maximal. This number is then
the total number of massless fermions in that representation.

  Locality of $R({\bf x},t)$ and boundedness of $\car^{-1}({\bf p})$
imply that $\car^{-1}({\bf p})$ is analytic except at generalized
degeneracy points. Furthermore, the allowed forms of singularities,
eqs.~(9) and~(11), imply that $\car^{-1}({\bf p})$ has a continuous
first derivative at the generalized degeneracy points, and that all
zeros of $\car^{-1}({\bf p})$, which occur at primary singularity
points, are of the relativistic form~(1). In addition,
$\car^{-1}({\bf p})$ is hermitian. Hence $\car^{-1}({\bf p})$ satisfies
all the assumptions of the Nielsen-Ninomiya theorem. Applying the
theorem, we conclude that $\car^{-1}({\bf p})$ has an equal number of
left handed and right handed zeros.
Since the set of interpolating fields
we have chosen is maximal, this implies that the spectrum contains
an equal number of left handed and right handed fermions in the given
complex representation.

\vspace{1ex}

  For completeness, we note that if a fermion belongs to a real
representation, it can generate both a left handed and a right handed
pole in its two point
function. This is the only way to violate the one-to-one correspondence
between poles of the two point function and massless fermions. Of course,
this exceptional situation is of no help if we are trying to construct
chiral fermions.

\vspace{2ex}

  Perhaps the most striking consequence of our theorem is the
constraints it puts on any attempt to
reproduce the standard model on the lattice without violating gauge
invariance. It asserts any such attempt must fail, if the
spectrum can be correctly determined by switching off the Electro-Weak
interactions, and provided that the effective vertex functions of
the Electro-Weak currents are pole-free in the symmetric phase.
In order to reach this conclusion there is no need to switch off QCD!
The reason is that the spectrum of QCD does not contain massless bosons,
and so, in the absence of the photon,
an effective lagrangian of the kind described above
is valid at distances larger than one Fermi. We comment
that it should be possible to accommodated a massless pion without
changing the conclusions because a Goldstone boson has only derivative
couplings.

  The fact that gauge invariant lattice theories are necessarily
vector-like raises an intriguing question concerning the relation between
fermion doubling and the anomaly. If we are careful to work with
an anomaly free theory and to break explicitly at the lattice scale
all global symmetries which are anomalous in the target continuum  theory
(\eg Baryon number), then there is no ``need''
for the appearance of doublers~[12]. This is the
case, for example, in the Eichten-Preskill model~[7].
The lattice theory does not have a bigger symmetry compared to the
target theory, and so one could expect to obtain the latter in the
continuum limit of the lattice theory. Nevertheless,
the doublers {\it do} appear. The resolution of this paradox, which is
clearly of a non-perturbative nature, must await
future investigations.

  I thank A.~Casher for numerous discussions of the subject.
This research was supported in part by the Basic Research Foundation
administered by the Israel Academy of Sciences and Humanities, and by
a grant from the United States -- Israel Binational Science Foundation.

\centerline{\rule{5cm}{.3mm}}

\newcounter{00001}
\begin{list}
{[~\arabic{00001}~]}{\usecounter{00001}
\labelwidth=1cm}

\item K.G.~Wilson, in {\it New Phenomena in Sub-Nuclear Physics} (Erice,
1975), ed. A.~Zichichi (Plenum, New York, 1977).

\item L.H.~Karsten and J.~Smit, \NPB{183} (1981) 103.

\item H.B.~Nielsen and M.~Ninomiya, \NPB{185} (1981) 20,
{\it Errata} \NPB{195} (1982) 541; \NPB{193} (1981) 173.

\item S.D.~Drell, M.~Weinstein and S.~Yankielowicz, \PRD{14} (1976) 487, 1627.
C.~Rebbi, \PLB{186} (1987) 200.

\item L.H.~Karsten and J.~Smit, \NPB{144} (1978) 536; \PLB{85} (1979) 100.

\item M.~Campostrini, G.~Curci and A.~Pelissetto, \PLB{193} (1987) 279.
A.~Pelissetto, \APH{182} (1988) 177.

\item J.~Smit, \APP{17} (1986) 531. P.~Swift, \PLB{145} (1984) 256.
E.~Eichten and J.~Preskill, \NPB{268} (1986) 179.
 I.~Montvay, \PLB{199} (1987) 89; \NPBP{4}  (1988) 443.

\item M.F.L.~Golterman, D.N.~Petcher and  J.~Smit, \NPB{370} (1992) 51.
W.~Bock, A.K.~De, C.~Frick, K.~Jansen and T.~Trappenberg,
\NPB{371} (1992) 683.
M.F.L.~Golterman, D.N.~Petcher and E.~Rivas, Wash. U. preprint HEP/92-80,
\NPB{ } to appear.

\item D.B.~Kaplan, \PLB{288} (1992) 342.

\item Y.~Shamir, in preparation.

\item N.~N.~Bogoliubov and D.~V.~Shirkov, {\it Introduction to the Theory
of Quantized Fields}, Interscience Publ. New York, 1959, p. 654.
H.~J.~Bremermann, R.~Oehme and J.~G.~Taylor, \PR{109} (1958) 2178.
S.G.~Krantz, {\it Function theory of Several Complex Variables}, John Wiley,
New York, 1982, p. 133.

\item T.~Banks and A.~Dabholkar, \PRD{46} (1992) 4016.

\end{list}

\end{document}